# Polarization-dependent all-optical modulator with ultra-high modulation depth based on a stereo graphene-microfiber structure

**Jin-hui Chen, Bi-cai Zheng, Guang-hao Shao, Shi-jun Ge, Fei Xu,* and Yan-qing Lu**

National Laboratory of Solid State Microstructures and College of Engineering and Applied Sciences, Nanjing University, Nanjing 210093, P. R. China

*feixu@nju.edu.cn

## Abstract:

We report an in-line polarization-dependent all-optical fiber modulator based on a stereo graphene-microfiber structure (GMF) by utilizing the lab-on-rod technique. Owing to the unique spring-like geometry, an ultra-long GMF interaction length can be achieved, and an ultra-high modulation depth (MD) of ~7.5 dB and a high modulation efficiency (ME) of ~0.2 dB/mW were demonstrated for one polarization state. The MD and ME are more than one order larger than those of other graphene-waveguide hybrid all-optical modulators. By further optimizing the transferring and cleaning process, the modulator can quickly switch between transparent and opaque states for both the two polarization states with a maximized MD of tens of decibels. This modulator is compatible with current fiber-optic communication systems and may be applied in the near future to meet the impending need for ultrafast optical signal processing.

Keywords：  graphene, polarization, microfiber, optical modulation

Two-dimensional materials, especially graphene, have attracted global interest [1], and many extraordinary electrical, mechanical, and optical properties have been reported. Regarding the exceptional optical properties, for example, 2.3% linear absorption (related to the fine structure constant) [2], saturable absorption, and a tunable Fermi level (either by electric gating, chemical doping, or even plasmon-induced doping [3]) have been demonstrated for a number of broadband optical devices based on graphene in both normal-incidence and waveguide-integrated configurations, such as mode-lock lasers [4,5], ultrafast photo-detectors [6,7], and modulators [8]. Among these, the optical modulator is a key device of optical communication that converts optical or electrical data into optical signals. High modulation speed, sufficient modulation depth (MD), and large optical bandwidth are indispensable for an optical modulator. Although the tuning of Fermi energy using electric gating on graphene to achieve modulators of gigahertz-level modulation speed has been demonstrated [8], it is difficult to further improve the modulation speed limited by the response time of the bias circuit. All-optical modulation can overcome the limitation on the modulation rate by using one light beam to control the transmission of a certain light beam. This type of modulation may avoid electrical-optical-electrical conversions, rendering the modulation process faster and less noisy. Several papers have reported all-optical modulation schemes with a high speed of 200 GHz based on graphene-microfiber-integrated devices by covering or wrapping a graphene sheet on a straight microfiber (MF). However, those schemes are either of low MD for a relatively short light-graphene interaction length of ~ 20μm, as in Ref [9], or of low modulation efficiency (ME) with the use of a thick microfiber of ~8-μm thickness, as in Ref [10]. The key issue for improving the performance of graphene-microfiber (GMF) all-optical modulator is to increase the optical interaction length. However, it is challenging to handle such a thin MF for graphene integration with sufficient lengths and strengths of interaction for practical applications.

Recently, on the basis of the rod-wrapping technique [11], we developed a reliable approach to fabricate GMF-integrated devices by wrapping an MF on a graphene-specialized rod. This method is much simpler than conventional integration techniques because we only need to coat a small piece of graphene on a thick rod, rather than a thin MF. Theoretically, with a strong evanescent field, the GMF interaction length can be arbitrarily increased using a spring shape of as many turns as is spatially possible. Owing to the extremely asymmetrical cross-section of such a stereo GMF device and nature of graphene-light interaction [8], it leads to different types of losses in the two fundamental modes in this hybrid structure. A broadband polarizer can be naturally integrated in such a device. We have demonstrated a broadband single-polarization resonator [12]. In the present study, we show that the stereo GMF structure can also operate as a polarization-dependent all-optical graphene modulator at around 1550 nm with a very high MD (~7.5 dB) and high ME (~0.2 dB/mW) that are more than one order larger than those reported previously [10]. In fact, the MD can be increased to tens of decibels and both of the two polarization states will be nearly fully transmitted at on state, if we can fabricate pure and clean GMF devices by improving the transferring and cleaning process. This type of fiber-integrated in-line all-optical modulators has great potential in fiber optical communication, in which there are demands for high-speed, wideband, low-cost, and integrated methods to modulate information.

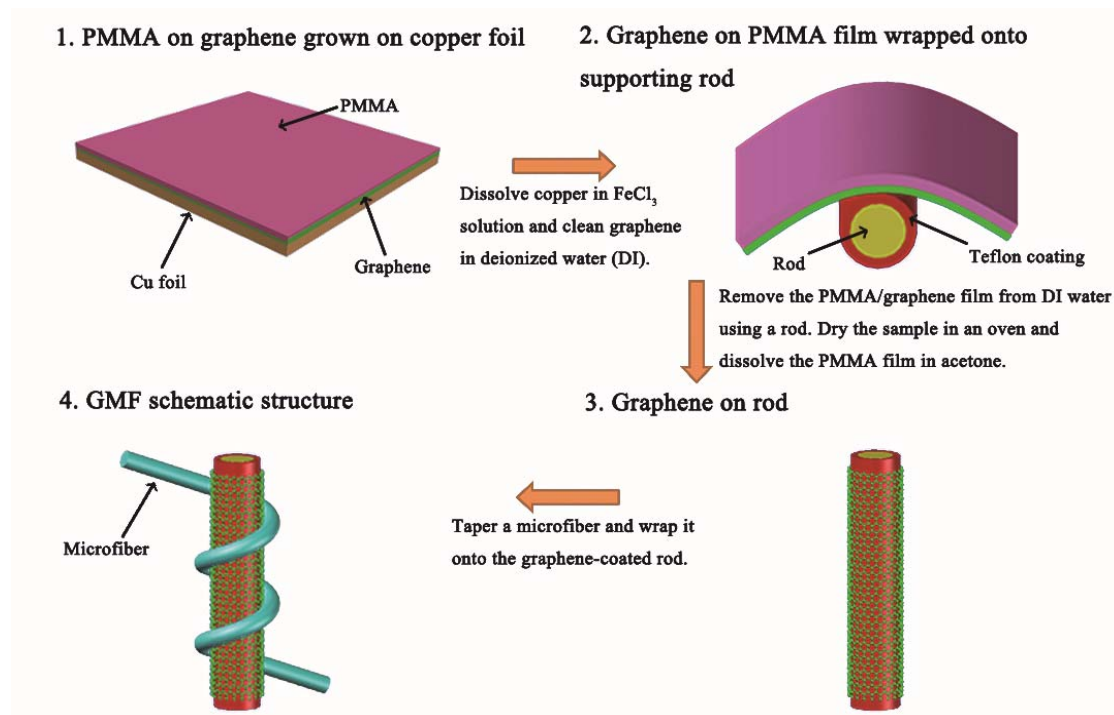

Figure 1. Fabrication process of the stereo hybrid GMF device. Panels 1-3 demonstrate the detailed steps for transferring graphene onto a rod. Panel 4 shows a schematic structure of the GMF device.

Figure 1 shows the fabrication process of the stereo GMF device. To prevent the loss induced by the relatively high index of the rod (polymethyl methacrylate or PMMA of diameter ~2 mm), a thin low-index Teflon layer (Teflon AF 601S1-100-6, DuPont, tens of micrometers in thickness with a refractive index of ∼1.31) was dip-coated on the rod's surface and dried in air for several hours. The graphene on Cu foil (Six Carbon Technology) was first spin-coated with PMMA film and dried it in oven for certain time. After dissolving the Cu foil and cleaning in deionized water (DI), the PMMA/graphene film was removed from DI water using a rod. The surface tension would lead to the film tightly encapsulating the rod when the film leaves DI water. The Raman spectrum clearly shows the single-layer nature of the transferred graphene [13], as shown in Fig. 2. In our experiment, we chose an MF with a diameter of ~3 μm, and there were two MF coils on a rod, which implied that the light-graphene interaction length was ~12 mm. And the MF in our experiment was tapered from conventional single mode fiber by utilizing flame-brushing technique

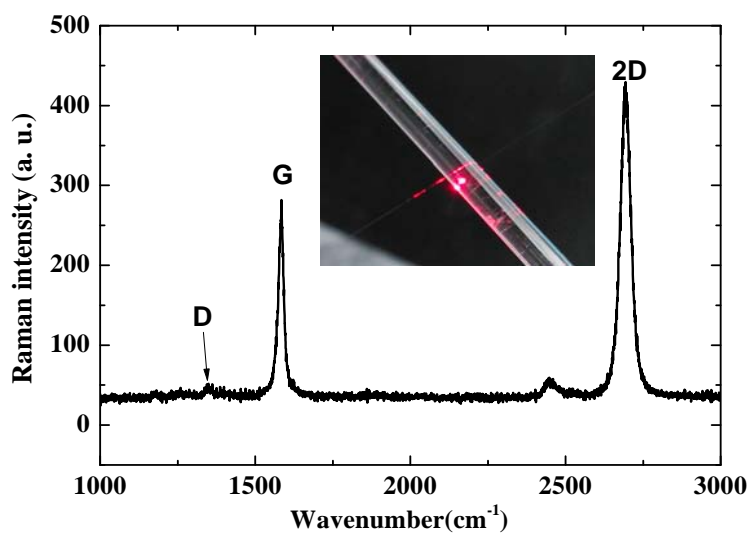

Figure 2. Raman spectrum of a monolayer graphene film. The Raman spectroscopy was carried out by using excitation sources: the 488 nm (2.54 eV) lines of an Ar ion laser. The laser beam was focused onto the sample by a 50X microscope objective lens (0.8 N.A.) and the scattered light was detected by a charge-coupled-device (CCD) detector. Inset, camera image of the as fabricated GMF device, launched with 650 nm light source. The strong scattering region indicates the graphene-MF interaction region.

The all-optical modulation mechanism is based on the Pauli blocking effect, as illustrated in Fig. 3a. When a high-power pump light at a short wavelength excites dense carriers in graphene from the valance band to the conduction band, the non-equilibrium carrier distribution broadens and equilibrates with the intrinsic carrier population through carrier-carrier scattering. The scattering process only lasts for hundreds of femtoseconds. Subsequent cooling and decay of the hot distribution through carrier-phonon scattering occurs on a time scale of picoseconds [14]. Carriers relaxing from the conduction band will lead to a band-filling effect, which results in a significant reduction of graphene absorption for light of greater wavelength. In this manner, we can use a high-power pump light source to switch a weak probe light source. As we can see above, the photon-excited carrier's relaxation process is dominated by carrier-phonon scattering, which lasts for several picoseconds. In other words, the all-optical modulation speed can theoretically reach hundreds of gigahertz. This has been experimentally demonstrated by using ultrafast optical pump-probe spectroscopy to measure the decay time of graphene carriers coated on a straight MF [9]. Moreover, the MD increases with the light-graphene interaction length and strength, which is much more enhanced in the stereo GMF platform than in other one-dimensional GMF platforms.

In our experiment, we first used a continuous-wave (CW) pump light source at 980 nm (OPEAK, Pump-LSB-980-500-SM) to measure the static modulation characteristics of our polarization-dependent stereo GMF platform, as shown in Figure 3b. Because the amplified spontaneous emission (ASE) light source (Connet C-ASE Optical Source) is non-polarized, a linear polarizer and a half-wave plate are incorporated between the light source and fiber wavelength-division multiplexer (WDM) to excite different eigen-modes of our GMF selectively with the same launching power. The ASE power input in our experiment was fixed as ~300μW. As shown in Fig. 3c, when the pump light is switched off, the output intensity difference between the low absorption-loss mode (LAM) and the high absorption-loss mode (HAM) is approximately 15 dB at 1550 nm, which is attributable to the broken symmetry of the GMF geometry and the absorption nature of graphene. The difference can be reductive and even eliminated by increasing the pump power. The small ripple superposed in the output spectra is attributable to the interface interference caused by the half-wave plate.

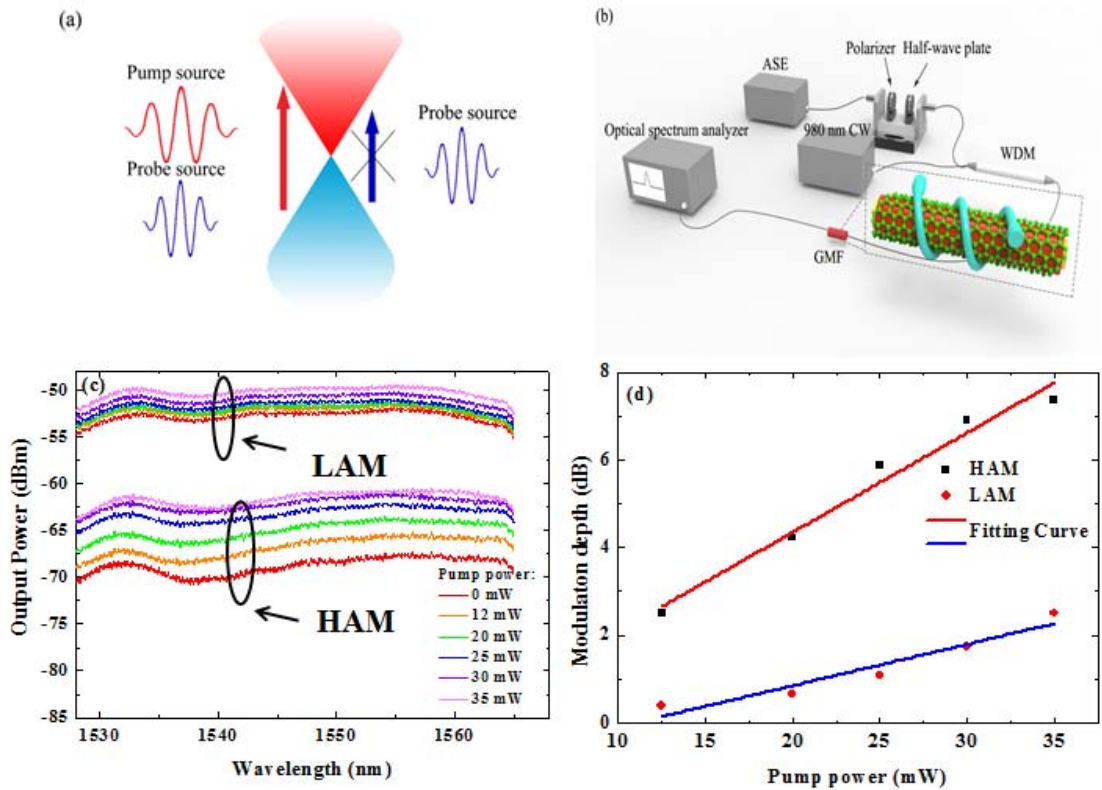

Figure 3. CW-pump-based modulation measurement system and results. (a) Schematic illustration of the all-optical modulation mechanism. (b) CW pump probe measurement system. (c) Output spectra of the stereo GMF device with a 980-nm pump source with different powers for the LAM and HAM. (d) Change in MD with the pump power for 1550-nm probe wavelength, extracted from (c). The blue and red lines are numerical fitting curves.

As the pump power increases, the output intensities of the HAM and LAM also increase because of the saturable absorption properties of graphene, as shown in Fig. 3c. Here, we define the MD as the transmission change of one polarization mode in the GMF device. The ME is the MD of the GMF device per unit pump power. In Fig. 3c, when the pump power reaches 35 mW, the MDs of the HAM and LAM are 7.5 dB and 2.5 dB, respectively, in the range of 1525 nm-1565 nm. Accordingly, the MEs of the HAM and LAM are ~0.2 dB/mW and 0.07 dB/mW, respectively. In Fig. 3d, we can see that the MDs of the GMF increase linearly with pump power. To further illustrate the relation between the MD and the pump power, we build a classical small-signal amplification model [15] to fit the experimental data (see Supplementary Information Note 1). It should be noted that the linear relation between the MD and the pump power holds only for the of small-signal condition. As the MD increases with the pump power, it will reach a saturation value, and the linear relation will naturally break down. In our numerical calculation, the theoretical maximized MD of our platform is ~23 dB and 5.4 dB for the HAM and LAM, respectively (see Supplementary Information Note 2) for a 3.0-μm-diameter MF with a two-coil structure. For a 2.5-μm-diameter MF, the MD of the HAM and LAM should be 42 dB and 9.6 dB (L ~ 12 mm), respectively. Regarding our ultra-long graphene-MF interaction length, the MD can be further improved if we optimize the coil number and MF diameter. Then, the HAM can switch between full transmission and full absorption with high-speed.

As we can see, our current experimental results only reach one third of theoretical values at 3.0 μm diameter. One of the reasons is that the maximized pump power we used did not saturate at the entire GMF interaction length. The reason why we did not increase the pump power further in our experiment was that there also existed a breakdown power threshold of the GMF because many residues of the protecting film of PMMA remained on the graphene surface after the PMMA removal process, which caused a great heating effect when launched with a high pump power. If the fabrication method is perfected (for example, the PMMA can be totally removed and only pure graphene will be left), the MD of our GMF can be further improved to approach the theoretical value. In order to observe the time-domain response of the GMF modulator, we also used a nano-second laser (KEOPSYS, PYFL-K04-RP-030-006-050-1064-T0-ET1-PK5D-FA) at 1064 nm to conduct the pump probe experiment. The pump probe system is illustrated in Fig. 4a. We used a bandpass filter centered at 1550 nm with a full width at half maximum (FWHM) of 12 nm to filter out the pump pulse laser. The modulated probe light then be was detected by a photo-detector (New Focus, 1544-B) and the transformed electric signal could be analyzed by an oscilloscope (Agilent Technologies, DSO-X 4024A). When the pulse laser (pulse width of ~6 ns, repetition rate of 100 kHz) of average power 1.5 mW was launched into the GMF device along with a CW 1550-nm probe light (Agilent, tunable laser 81980 A) of power ~5 mW, the pulse laser increased the probe-light transmission of the GMF device in their presence and switched off the probe source when they were absent. This can be readily seen in Fig. 4b. When the probe laser was switched off, there was no modulated serial signal, which reflects the excellent filtering of the pump light. When the probe laser was turned on, a modulated probe light was clearly observed, which unambiguously demonstrated that our GMF device could serve as an all-optical modulator. It should be noted that although we only demonstrated 100-kHz modulation repetition, the GMF structure can be operated at sub-THz repetition rates for the ultrafast carrier relaxation of graphene [9].

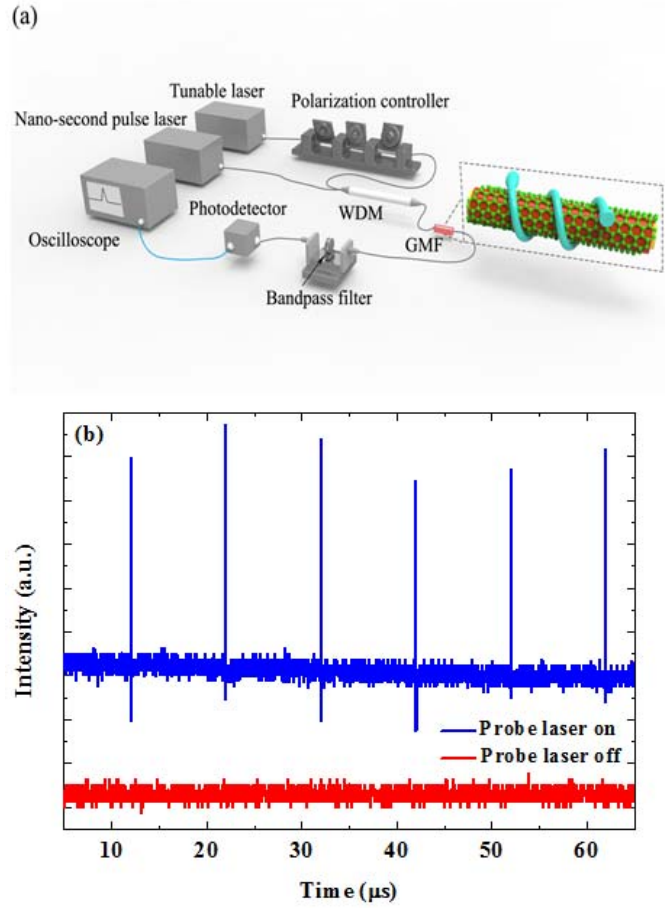

Figure 4. Pulse-laser pump probe system. (a) The optical measurement system. (b) The time-domain response of the GMF modulator. The modulated probe light curve was moved up for clarity.

In conclusion, on the basis of a GMF-integrated stereo device, we demonstrated a highly polarization-sensitive all-optical modulating platform. The unique geometry practically enabled sufficient light–graphene interaction length and strength, and a very high MD of ~7.5 dB and a high ME of ~0.2 dB/mW were simultaneously achieved for the HAM. The polarization-dependent modulation characteristics and time-domain response were also investigated. The performance was limited by the relatively low quality of the graphene used in the platform. An MD as high as the theoretical prediction (~ 23 dB at 3μm diameter) can be expected by improving the transferring and wrapping process of graphene. The findings of this study may contribute to many new future applications in all-optical fiber-optic circuits.

ASSOCIATED CONTENT
Supporting Information
Section 1: Small signal amplification model. Section 2: Numerical simulations and results.

Notes
The authors declare no competing financial interest.

ACKNOWLEDGEMENTS

The authors thank Hao Qian and Song-hua Cai for their assistance on drawing the schemes. This work is sponsored by National 973 program (2012CB921803, 2011CBA00205), National Science Fund for Excellent Young Scientists Fund (61322503), National Science Fund for Distinguished Young Scholars (61225026) and the Fundamental Research Funds for the Central Universities.